# Fine pitch CdTe-based Hard-X-ray polarimeter performance for space science in the 70-300 keV energy range

S. Antier[a1], O. Limousin[a], P. Ferrando[a]

[a]CEA-Irfu, CEA Saclay, F-91191 Gif-sur-Yvette Cedex, France.
[1]Corresponding author e-mail: sarah.antier@cea.fr

**Abstract**   X-rays astrophysical sources have been almost exclusively characterized through imaging, spectroscopy and timing analysis. Nevertheless, more observational parameters are needed because some radiation mechanisms present in neutrons stars, black holes or AGNs are still unclear. Polarization measurements will play a key role in discrimination between different X-ray emission models. Such a capability becomes a mandatory requirement for the next generation of high-energy space proposals. We have developed a CdTe-based fine-pitch imaging spectrometer, Caliste, able to respond to these new requirements. With a 580-micron pitch and 1 keV energy resolution at 60 keV, we are able to accurately reconstruct the polarization angle and polarization fraction of an impinging flux of photons which are scattered by 90° after Compton diffusion within the crystal. Thanks to its high performance in both imaging and spectrometry, Caliste turns out to be a powerful device for high-energy polarimetry. In this paper, we present the principles and the results obtained for this kind of measurements: on one hand, we describe the simulation tool we have developed to predict the polarization performances in the 50-300 keV energy range. On the other hand, we compare simulation results with experimental data taken at ESRF ID15A (European Synchrotron Radiation Facility) using a mono-energetic polarized beam tuned between 35 and 300 keV. We show that it is possible with this detector to determine with high precision the polarization parameters (direction and fraction) for different irradiation conditions. Applying a judicious energy selection to our data set, we reach a remarkable sensitivity level characterized by an optimum Quality Factor of 0.78 ± 0.06 in the 200-300 keV range. We also evaluate the sensitivity of our device at 70 keV, where hard X-ray mirrors are already available; the measured Q factor is 0.64 ± 0.04 at 70 keV.

___________________________________________

## 1. Introduction

Polarization measurement provides new information about magnetic field structure, emission process and geometry in several X-ray sources, like accreting black holes in stellar mass systems, AGNs, and neutron stars. At the moment, X-ray studies are still mainly based on imaging, spectral and timing analysis. Apart from GAP onboard the IKAROS satellite (Yonetoku 2011)[1], no dedicated polarimeters have been launched so far. The SPI and IBIS telescopes onboard INTEGRAL recently measured polarized emissions from the Crab pulsar, Cygn X-1(Laurent et al 2014)[2] and Gamma Ray Burst (Götz et al 2014)[3]. The growing of interest in polarimetry and the progress in focusing optics in the hard X-ray domain leads to new mission proposals such e.g. COSPIX (Ferrando et al. 2010)[4] or PheniX (Roques et al. 2012)[5]. Within the next decade, only two X-ray missions will be able to measure linear X-ray polarization (with Compton properties): NuSTAR (Lotti et al 2013)[6] and Astro-H (Tajima et al. 2010) [7]. On Astro-H, the SGD detector is sensitive to polarization in the range of 50 – 200 keV with a 30 $cm^2$ effective area whereas the NuSTAR detectors are sensitive in the 3-80 keV energy range with 80 $cm^2$. Within this scientific frame, we propose a Compton polarimeter based on the Caliste technology: a fine pitch, high energy and good time resolution imaging spectrometer developed at CEA (Limousin et al. 2011)[8]. In this paper, we quantitatively investigate the polarimetric performance that can be achieved with such a detector in the 70-300 keV energy range, both from measurements taken with a linearly polarized beam at the European Synchrotron Radiation Facility (ESRF) in Grenoble (France), and from simulations dedicated to this detector configuration.

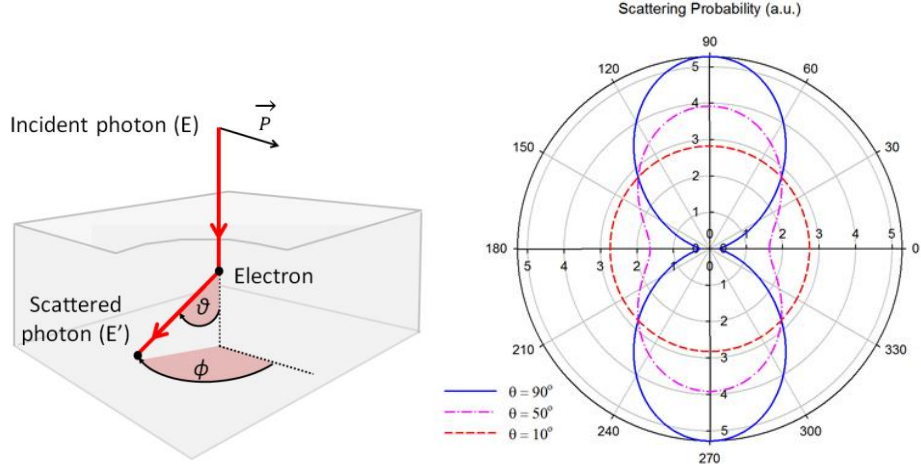

Figure 1 (a) Scheme of Compton scattering of a polarized photon; (b) Azimuthal angle ($\varphi$) probability distribution for a given Compton scattering angle ($\theta$) of linearly polarized photons at $E = 200$ keV. The direction of the polarization is parallel to the horizontal axis of the polar plot.

## 2. Scattering Polarimetry principle

In the hard X-ray domain, linear polarization with spectro-imaging detectors is possibly performed using Compton properties by especially studying the anisotropic azimuthal distribution of the scattered photons as shown on Figure 1. In the detection plane and with a polarized beam, Klein-Nishina (1929)[9] describes the differential cross-section for Compton scattering per unit of elementary solid angle $d\Omega$:

$$\frac{d\sigma}{d\Omega} = \frac{r_0^2}{2}\left(\frac{E'}{E}\right)^2 \left[\frac{E'}{E} + \frac{E}{E'} - 2\sin^2\theta\cos^2\varphi\right] \quad (1)$$

where $r_0$ is the classical electron radius, $E, E'$ are respectively energies of incoming and scattered photons and $\varphi$ is the azimuthal deviation angle formed by the scattering plane and the incoming photon polarization plane. The secondary photon is scattered from its original direction by $\theta$. $E$, $E'$ and $\theta$ are linked by the following equation:

$$E' = \frac{E}{1 + \frac{E}{m_e c^2}(1 - \cos\theta)} \quad (2)$$

with $m_e c^2$ being the electron rest energy.

The asymmetry of the scattered photons distribution increases with the scattering angle $\theta$ and reaches its maximum at $\theta=90°$. Using a single detector, the best configuration to study spatially this distribution is to select only 90° scattered photons using the relation between energy and scattered angle (Equation 2). In this case, the spatial distribution of scattered photons is very asymmetric and looks like a bowtie. This kind of measurements implies a detector able to record multiple events (Compton), with an excellent energy resolution (mandatory for 90° scattered photons selection) and a very small pixel size (spatial distribution). The sensitivity of a polarimeter is expressed by the figure of merit and especially the polarimetric modulation factor $Q$. For a pixelated detector, $Q$ is written as:

$$Q = \frac{Nmax - Nmin}{Nmax + Nmin} \quad (3)$$

where $Nmax$ and $Nmin$, are the maximum and minimum of the angular azimuthal distribution of the scattered photons defined over the detector plane. Because of the nature of the scattering process, $Nmax$ and $Nmin$ are counted along two orthogonal directions.

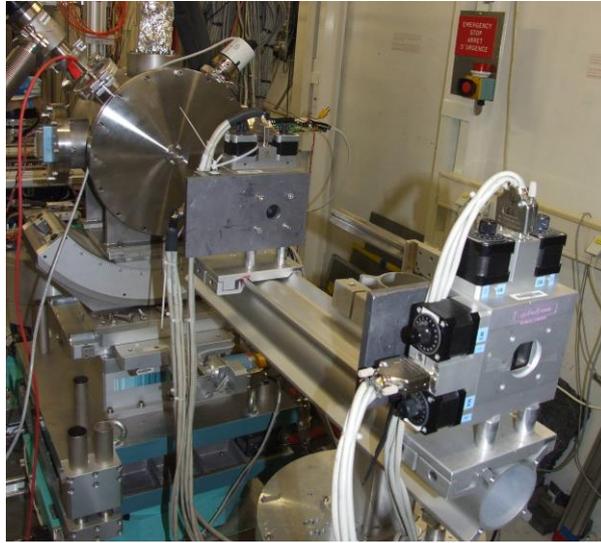

Figure 2. Picture of the PolCaliste experiment set-up at the experimental hutch of the ESRF ID 15A beam line in Grenoble (France).

## 3. Experimental Set-up

Caliste-256, a set of CdTe-based fine-pitch imaging spectrometers, responds to the requirements needed for good polarimetric performance. One Caliste device consists of a 1 cm$^2$ pixelated CdTe/CZT detector (1or 2 mm thick) mounted on top of a 4-side buttable hybrid module supporting dedicated low-noise front-end electronics. It is able to detect photons between 2 keV and 250 keV and to measure simultaneously energy deposits in different pixels. With a 580-micron pitch, a 16x16 pixels array and an achievable 1 keV energy resolution at 60 keV, Caliste-256 is specially designed for various hard X-ray and gamma-ray space missions. This small module turns out to be a powerful tool for high-energy polarimetry; it was designed to be able to accurately reconstruct the polarization angle and polarization fraction of an impinging flux of photons that have Compton scattered by 90° in the crystal.

The experimental polarimetric performances of Caliste-256 have been tested with the high-energy and mono-energetic ID15 beam line at ESRF (Grenoble, France). During the acquisition, the beam was directed onto the top surface of the detector - confined in a vacuum chamber - through a collimation system, as shown on Figure 2. The beam shape, set to 100 μm × 100 μm, is smaller than the 500 μm × 500 μm pixel size. A scan of the entire detector surface provides a fine energy calibration of each individual pixel. The detector response uniformity was also checked. For runs dedicated to polarization studies, the beam spot was centered onto one of the four central pixels and the detector plane was maintained perpendicular to the beam axis. Several measurements have been performed for both Caliste samples (1 mm CdTe and 2 mm CZT) for different operating conditions: with beam energies from 70 to 300 keV, degree of polarization from 80 to 98% and angle of polarization between 0 and 30°.

## 4. Simulation tool of the polarimetric performance

A 3D numerical simulation tool has been developed to predict Caliste-256 polarization performances in the 50-300 keV energy range. It takes into account Caliste design (crystal composition, thickness and geometry) and irradiation conditions (incident beam energy and position, inclination relative to the detector surface, angle and degree of polarization).

Based on matter interactions probabilities, the model demonstrates the possible study of polarization through the absorption and detection of a scattered photon in a pixel. It also reproduces the theoretical azimuthal distribution of scattered photons. This model does not take into account more complex phenomena for which a full Monte-Carlo simulation is required (as e.g. the Doppler broadening of the Compton interaction or the measured energy dispersion of the detector). However, its complexity level is sufficient to describe our observations very well.

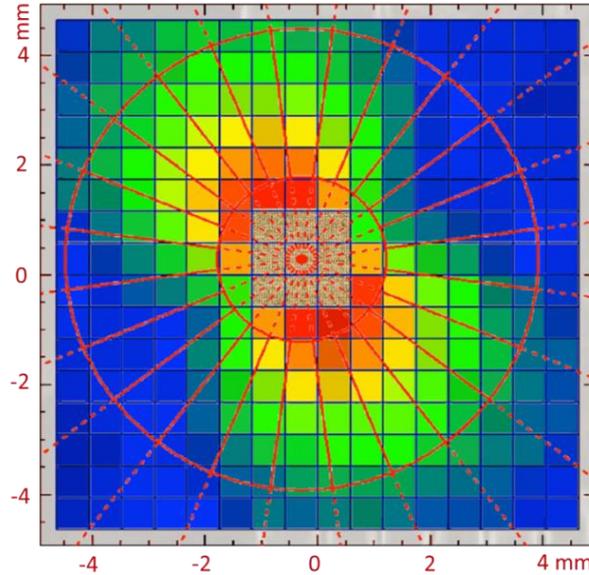

Figure 3. A false color map of the CdTe Caliste-256 counts distribution of events with scattered energies in the 139-148 keV band. These data are for a 200 keV photon beam, 98% linearly polarized and with 30° polarization angle. The 9 central pixels are off scale. On top of the data are shown the sectors, with their inner and outer radii, and their angular limits, which are used to count the events for building the azimuthal distribution curves.

### 5. Polarization measurement procedure

A judicious energy selection has been applied to our experimental dataset in order to select 90° scattered photons to maximize the modulation factor Q (compared to a lower efficiency). It takes also into account the Doppler broadening (Zoglauer and Kanbach 2003)[10] and the detector energy resolution (1 keV FWHM). Moreover, it limits the background noise resulting from charge sharing, fluorescence and backscattering of the photons scattered in the front-end electronics situated below the detector. It corresponds to an energy selection of 61 keV ± 2 keV at 69.5 keV incident beam energy, 144 keV ± 4 keV at 200 keV and 189 keV ± 8 keV at 300 keV. A 2D Compton event distribution map onto the detector surface shows the expected bowtie shape as shown on Figure 3, obtained with a 200 keV incident energy with a beam spot centered onto one of the central pixels.

Data recorded with a 69.5 keV incident energy correspond to the scan run. In this case, one difficulty is to reconstruct the bowtie as shown on Figure 3, as if the beam was centered to one of the central pixels. During the acquisition, the beam scans the total detection plane with a 4s stop in the center of each pixel. We isolate the 4s time range and we identify the pixel where the beam impinges by measuring the maximal number of interactions in a pixel. We superimpose each corresponding Compton event distribution map in order to agree the pixel of the beam spot, to one of the four central pixels. Some 4s acquisition relative to the border pixels, defining a corona of 4 pixels wide, are skipped because they bring poor polarization information at 70 keV incident energy. The 2D Compton event distribution reconstructed map with 69.5 keV incident beam energy, is shown on Figure 4.

The next step of the procedure consists in extracting the distribution of the Compton scattered photon count-rate with respect to the azimuthal angle. To do so, we superimpose a grid with 24 angular sectors (12 angular sectors for 70 keV data due to smaller scattering length relative to pixel size), of 15° each (30° for 70 keV), on the Compton event distribution map. The center of the grid coincides with the beam spot position on the central pixel. The inner and outer radii of the grid are set to 1.2 / 1.5 and 3.0 / 4.2 mm respectively for the 70/200-300 keV incident beam energy. The center of the grid coincides with the beam spot position on the central pixel. For each sector, the number of counts from a given pixel assigned to each sector is calculated by adding pixels which are fully included into the sector surface. If only a part of the pixel is included in the sector, then the number of counts, fractional, assigned to that sector is given by the total number of counts of the pixel weighted by the fraction of the pixel surface which intersects the sector. This procedure permits the construction of a smooth modulation curve as a function of angle, which gives access to polarization parameters. The same method is applied to the data obtained from the simulation and two comparable modulation curves are obtained.

Finally, for the determination of angle and fraction of polarization, we employed a fast fitting model with the "curvefit" function of IDL, applied to the modulation curve.

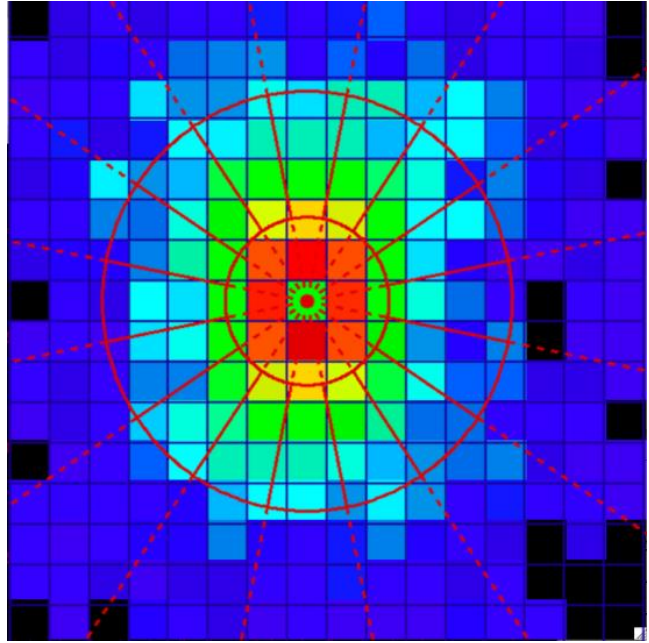

Figure 4. A false color map of the CdTe Caliste-256 counts distribution of events with scattered energies in the 60-62 keV band. These data are for a 69.5 keV photon beam, 98% linearly polarized and with 0° polarization angle.

### 6.  Caliste Polarization Performance in 200-300 keV energy range

From each experimental acquisition and the corresponding simulation, modulation curves using an energy selection of 90°-scattered photons have been built in order to estimate the sensitivity of Caliste and the polarization parameters have been measured (see chapter 5). The Q factor obtained is found to be 0.75 ± 0.04, similar to simulation results at different energies [200/300 keV], with different crystal compositions and thicknesses [CdTe/CZT], with a different degree of polarization [80-98% beam polarized] and with different angles of polarization [0-30°] (Antier et al (2014)[11]. When compared with other CdTe-based planar polarimeters, we obtain a better polarization Q factor (Curado da Silva et al. 2008)[12]. This is explained by the small lateral pixel size of the Caliste prototypes, which allows better modulation resolution (Caroli et al. 2005)[13], and the excellent energy resolution which allows an optimal angular selection of Compton double events.

Thanks to the fit of the angle of polarization, very good linear correlation between the incident beam polarization direction $\varphi_{beam}$ and the measured polarization direction $\varphi_{obs}$ has been obtained : $\varphi_{obs}$ = -0.5 (± 0.5) + *1.00 (±0.04)*  $\varphi_{beam}$. We reached similar results with Caliste degree accuracy at less than 5% for all our acquisition which is an acceptable performance for studying the different physical processes that generate polarized emissions levels. As an example, Figure 5 shows the experimental and simulated modulation curves for an expected angle of polarization of 0° and a fraction of polarization of 98%. The measured Q factor is 0.73 ± 0.02, the angle of polarization given by the fitting procedures is -0.6 ± 0.09 and the degree of polarization is 86% ± 0.2%. When compared to experimental data, the simulated data are normalized so that the total number of counts is the same for both sets. The model almost perfectly matches the data. In the simulation, we take into consideration the small inclination between the detector surface and the beam axis. Small differences between the measured angle and degree and beam angle and degree have been explained on one hand by this slightly inclination of the detector regarding the beam axis, and on the other hand the small background not identified in the experimental measurement, as well as the fact we did not take into account the Doppler broadening and energy resolution in the simulations.

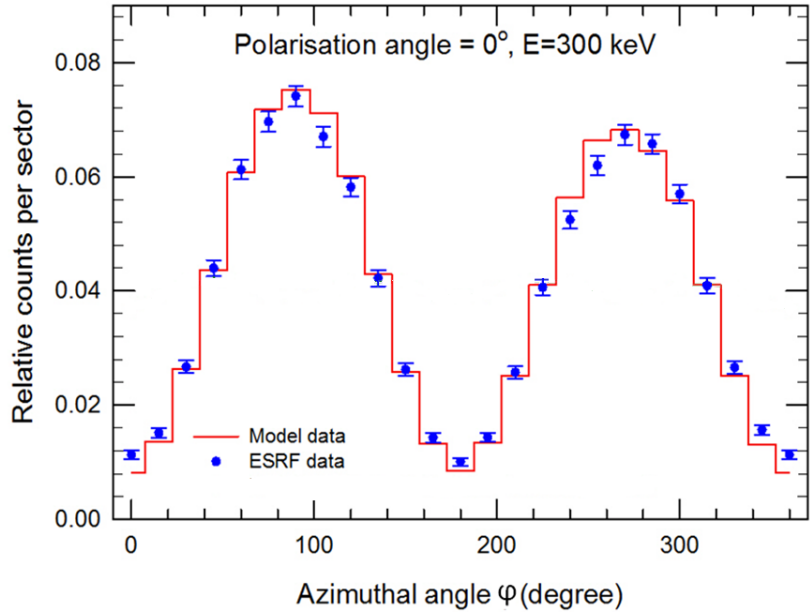

Figure 5. Angular azimuthal scattered photon distribution for two different angles of polarization, a 300 keV incident beam with a 98% linear polarization at an angle of 0°. A simulated modulation curve is superimposed to the data, taking into account the true incidence angle of the beam

### 7. Caliste Polarization Performance at 69.5 keV beam energy

High-energy performance results in terms of polarization have been obtained in experimental laboratory conditions. We expect poorer outcomes in space due to a higher background noise level. Nevertheless, not to devalue the polarimetry sensitivity, we need to collect a maximum number of photons, and especially photons which have been scattered with a 90° scattering angle. One issue is to use the Caliste as an elementary element of a detector plane. In this case, it would be better to employ the latest version of Caliste named Caliste HD, which has the advantage of having a power consumption 4 times lower compared to Caliste-256 and extends the dynamic range from 250 keV to 1 MeV with improved spectrometric performances (Meuris et al.)[14].

Another possibility would be to use Caliste disposed in the focal plane of X-ray focusing mirrors. Multilayers or coating-mirror technologies are currently in progress [15]. At the moment, multilayer optical mirrors are efficient up to 80 keV and are used for the NuSTAR mission, which has focal plane detectors similar to the Caliste technology (Rana et al.)[16]. Using the 69.5 keV-scan run of the Caliste detector acquired at ESRF, we reconstruct the 2D event map as if the beam was directed to one of the central pixels (with 0° angle and 98 % degree of polarization). Figure 6 represents the corresponding modulation curve in blue and the associated best fit in red. At 69.5 keV, the measured Q factor is 0.64 ± 0.04 and the best fitted polarization parameters attest an accuracy of 15 % regarding the degree of polarization and of 1° regarding the angle of polarization. These high polarization performances for such low energies demonstrate the possible study of polarization at energies below 80 keV, where hard X-ray mirrors are already available.

### 8. Summary

In this paper, we demonstrate the excellent potential of Caliste-256 (CZT and CdTe) for being used as an X-ray polarimeter in the 70-300 keV energy range. Thank to measurements taken with a linearly polarized beam at the European Synchrotron Radiation Facility (ESRF) in Grenoble, we investigate the polarimetric performances of Caliste in terms of sensitivity (modulation Q factor) and polarization parameters measurements (degree and angle of polarization accuracy). We performed 3D analytical simulations to support the experimental results: they faithfully reproduce the experimental conditions and Caliste design. The high energy and good time resolution permits to select Compton events scattering close to 90° where polarization information reaches its maximum. The results obtained show that Caliste detectors have a high sensitivity for polarization measurements in the energy range where hard X-ray mirrors are still efficient: the optimum quality Q factor is 0.64 ± 0.04 at 70 keV (and 0.78 ± 0.02 in the 200-300 keV range). We have also been able to measure the beam polarization parameters with an accuracy better than 1° for the polarization angle and better than 15% for the fraction of polarization at 70 keV beam energy (5% for 200-300 keV energy range). This demonstrates the capability with this Caliste detector to perform Compton-polarization measurements in a large energy range, of astrophysical bright sources in the context of space missions.

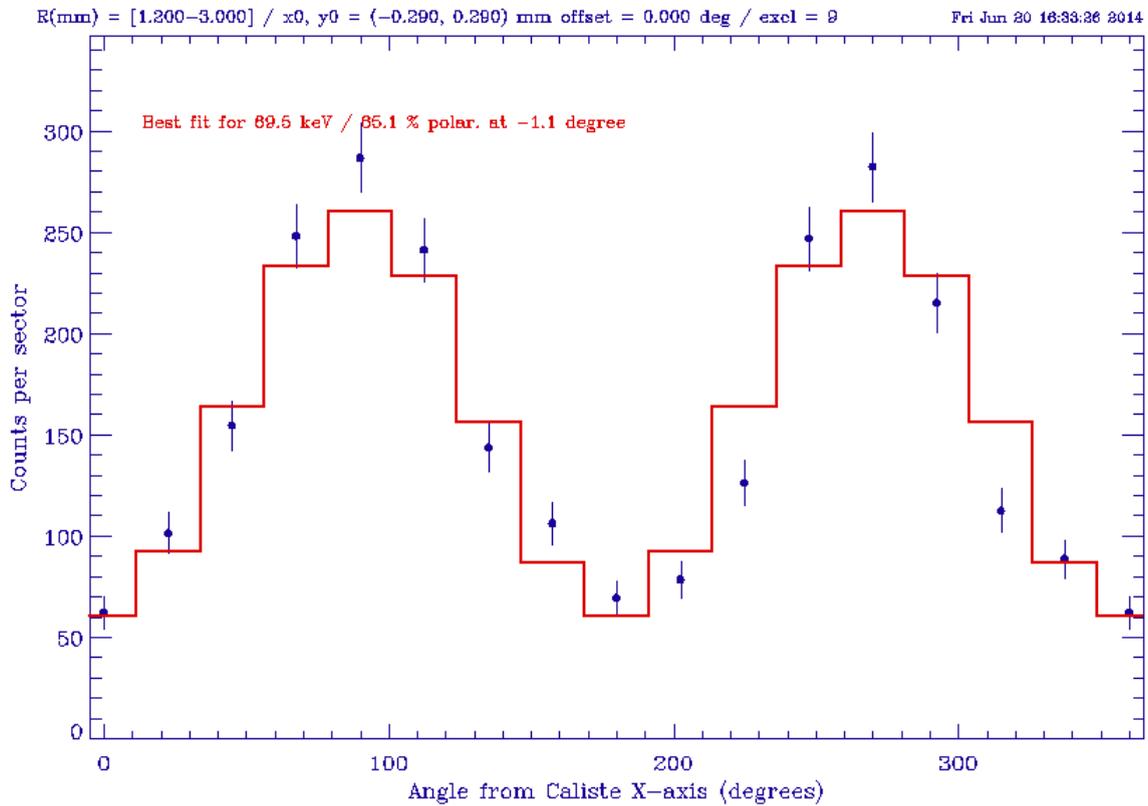

Figure 6. Angular azimuthal scattered photon distribution for a 69.5 keV incident beam with 98% linear polarization and an angle of 0°. Similar good fit in red with best fits of 85.8% polarization at -1.1° and 86.0 % at 28.6° respectively

**Acknowledgment**

The authors would like to thank E. Caroli, R.M. Curado da Silva, C. Blondel, R. Chipaux, V. Honkimaki, B. Horeau, P. Laurent, J.M. Maia, A. Meuris, S.Del Sordo and J.B. Stephen.